\documentclass[twocolumn,showpacs,preprintnumbers,amsmath,amssymb]{revtex4}
\usepackage{graphicx}
\usepackage{dcolumn}
\usepackage{bm}
\begin{document}    

\title{Apollonian networks}
\author{Jos\'e S. Andrade Jr.}
\author{Hans J. Herrmann}
\altaffiliation[Also at ] {Institute for Computer Physics 1, 
University of Stuttgart\\
email: hans@ica1.uni-stuttgart.de }
\affiliation{%
Departamento de F\'{\i}sica, Universidade Federal do Cear\'a,\\
60451-970 Fortaleza, Cear\'a, Brazil.}
\author{Roberto F. S. Andrade}
\affiliation{
Departamento de F\'{\i}sica, Universidade Federal da Bahia,\\
40210 Salvador, Bahia, Brazil.
}%
\author{Luciano R. da Silva}  
\affiliation{
Departamento de F\'{\i}sica, Universidade Federal do Rio Grande do Norte,\\ 
59072-970 Natal, Rio Grande do Norte, Brazil.  
}%

\date{\today}


\begin{abstract}
We introduce a new family of networks, the Apollonian networks, that
are simultaneously scale-free, small world, Euclidean, space-filling
and matching graphs. These networks have a wide range of applications
ranging from the description of force chains in polydisperse granular
packings and geometry of fully fragmented porous media, to
hierarchical road systems and area-covering electrical supply
networks. Some of the properties of these networks, namely, the
connectivity exponent, the clustering coefficient, and the shortest
path are calculated and found to be particularly rich. The
percolation, the electrical conduction and the Ising models on such
networks are also studied and found to be quite peculiar. Consequences
for applications are also discussed.
\end{abstract}

\pacs{89.75.Hc, 89.75.Da, 89.20.Hh, 64.60.Ak, 02.50.Cw, 05.50.+q}

\maketitle

The study of networks has fruitfully inspired the understanding of
transport and information flow within systems of many degrees of
freedom during the last years \cite{Barabasi02,Doro1,Watts,Vespignani}.  
Many network types have been proposed and investigated serving to
differentiate and elucidate effects ranging from clan-formation,
epidemics spreading and logistic planning to earthquake prediction,
neural activity and immunological defenses
\cite{Barabasi02,Doro1,Watts,Vespignani}.

Among a multitude of models, much attention has been dedicated
recently to the study of scale-free networks, namely, networks
that display a power-law degree distribution, $p(k) \propto
k^{-\gamma}$, where $k$ is connectivity (degree) \cite{Barabasi02}.
The widespread appearance of networks with fat-tailed degree
distribution in Nature has been mainly justified by the fact that
their evolution is self-organized in a way that hub elements are
naturally generated and represent the dominant constituents of the
network \cite{Doro3}.

There are, however, several conceptual and applied aspects of
scale-free network models that are still under investigation. For
instance, the fact that networks with exponent $\gamma > 2$ can be
embedded in regular Euclidean lattices has motivated Rozenfeld {\it et
at.} \cite{Rozenfeld02} to develop a method for generating random
scale-free networks with well defined geographical properties. The
{\it ultrasmall} characteristic found for the diameter of random
scale-free networks \cite{Cohen03} is also an intriguing aspect of
these structures that is expected to explain anomalies in their
diffusional and transport phenomena behavior. Finally, the stability
to crashes due to removal of sites in such networks has also been the
focus of intense research activity \cite{Cohen00}. These studies
revealed that random scale-free networks remain connected even if
nearly all of their nodes break down, as long as the degree
exponent $\gamma \leq 3$. As a consequence of this striking property,
the Internet and other well known networks displaying scale-free 
behavior must be resilient to random breakdown.  

An important issue that has been much less explored in network science
is the design problem of a suitable complex network for a given system
where connectivity plays an essential role. For this purpose, {\it
deterministic} networks represent the ideal strategy, since they are
controllable and can therefore be subjected to the particular
constraints of the problem, for example, those related to space, time,
economical and other types of limitations. Using deterministic models,
it should be possible to outline some specific guidelines for the
design of real networks with desired properties (e.g., high
clustering, compactness, etc.). Some examples of deterministic models
that are scale-free have been recently proposed and successfully used
to describe random growing networks \cite{Kahng01,Doro2}, but they
have eluded the possibility of being embedded in Euclidean space. In
this letter we introduce a class of networks, named Apollonian networks,
that can be either deterministic or random, are scale-free, display
small world effect, can be embedded in an Euclidean lattice, and show
space-filling as well as matching graph properties.

\begin{figure}
\begin{center}
\includegraphics[width=0.2\textwidth]{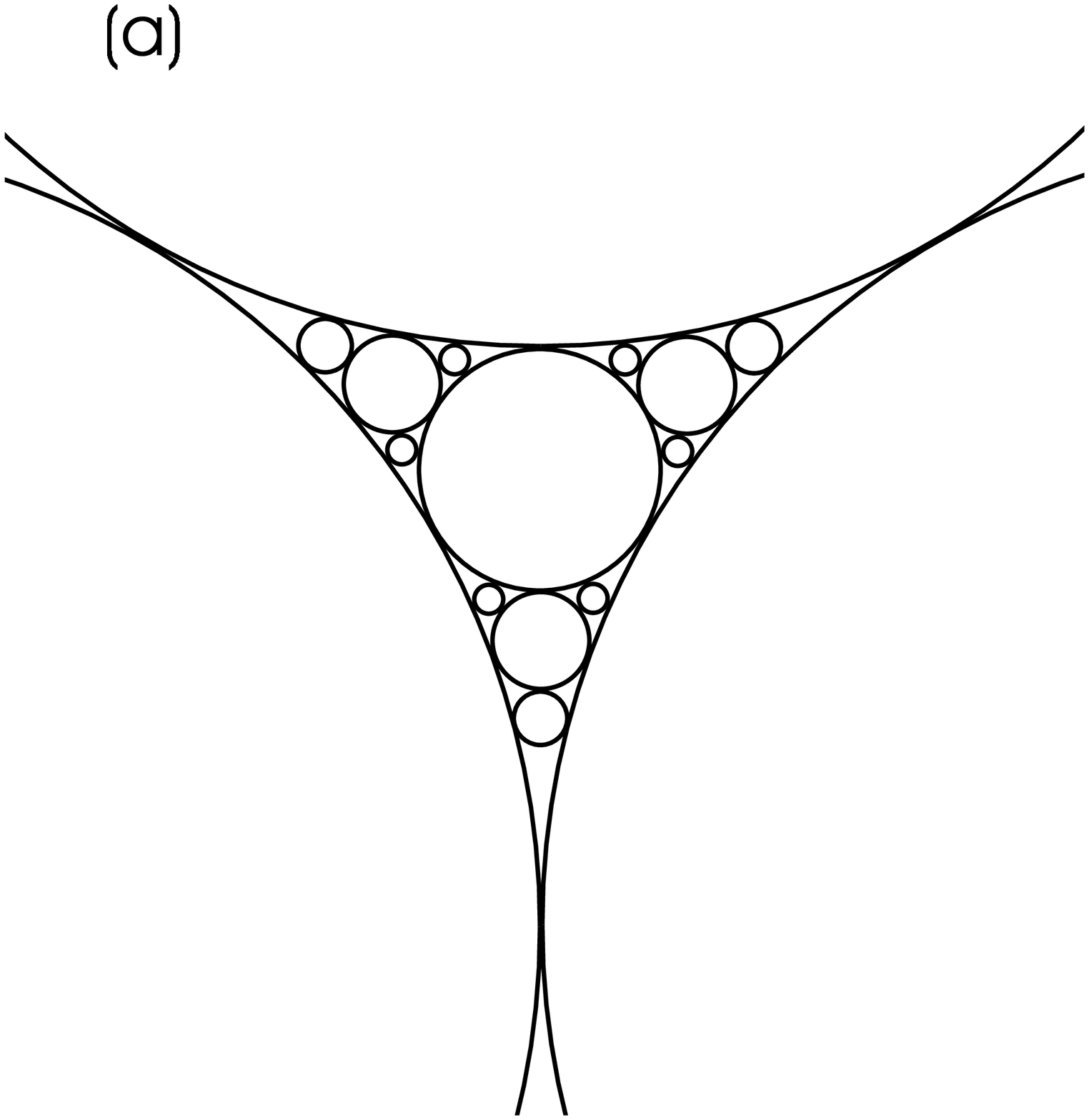} \
\includegraphics[width=0.24\textwidth]{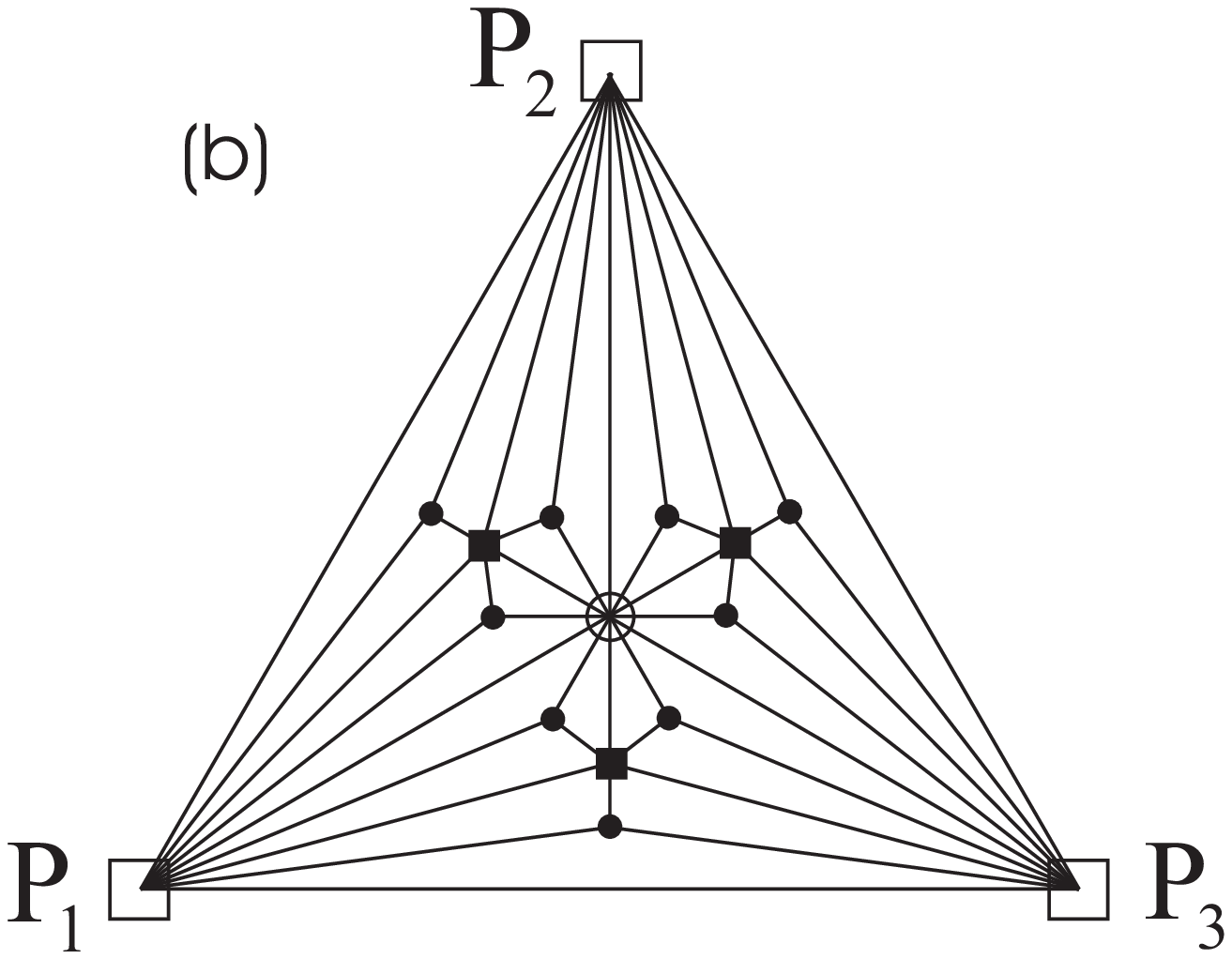} 
\end{center}
\caption[kurzform]{\label{Fig1} (a) Classical Apollonian packing. (b)
Apollonian network; first, second and third generation are symbols
$\bigcirc$, $\blacksquare$ and $\bullet$ respectively.}
\end{figure}

To define our network in its simplest deterministic version let us
start with the problem of space-filling packing of spheres according
to the ancient Greek mathematician Apollonius of Perga \cite{Boyd73}. 
In its classical solution, three circles touch each other and the hole
between them is filled by the circle that touches all three, forming
again three but much smaller holes that are then filled again each in
the same way as shown in Fig.~\ref{Fig1}(a). The circle size
distribution follows a power-law with exponent of about 1.3
\cite{Boyd73}. In fact there exist many other topologies with
different fractal dimensions \cite{Herrmann90}. This procedure can
also be generalized to higher dimensions \cite{Mahmoodi04}.

Apollonian tilings can describe dense granular packings and have also
been used as toy models for turbulence and fragmentation. Connecting
the centers of touching spheres by lines one obtains a network which
in the classical case of Fig.~\ref{Fig1}(a) gives a triangulation that
physically corresponds to the force network of the packing. We will
call the resulting network shown in Fig.~\ref{Fig1}(b) an ``Apollonian
network''. Besides resembling the graphs introduced by Dodds 
\cite{Dodds80} for the case of random packings, this network has
also been used in the context of porous media \cite{Adler85}.

\begin{figure}
\begin{center}
\includegraphics[width=0.4\textwidth]{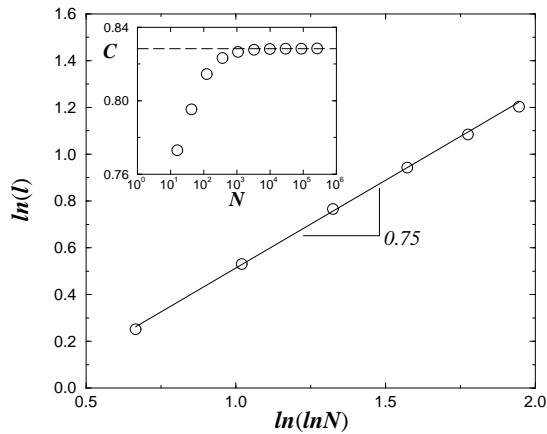}
\end{center}
\caption{ \label{Fig2} Shortest path $l$ as a function of the 
number of sites $N$. The solid line is a guide to the eye with slope
$3/4$. The inset shows the semi-log plot of the clustering coefficient
$C$ as a function of $N$. The dashed line is a guide to the eye for
$C=0.828$, observed at large values of $N$.}
\end{figure}

First we show that Apollonian networks are scale-free. From
Fig.~\ref{Fig1}(b) one can see that at each generation $n$ the number
of sites $N$ increases by a factor three and the coordination of each
site by a factor two. More precisely, at generation $n$ ($n=0$, $1$,
$2$,...) there are $m(k,n)=3^{n}$, $3^{n-1}$, $3^{n-2}$..., $3^{2}$,
$3$, $1$, and $3$ vertices with degree $k=3$, $3 \times 2$, $3
\times 2^{2}$,..., $3\times 2^{n-1}$, $3 \times 2^{n}$, and $2^{n+1}$,
respectively, where the last number of vertices and degree correspond
to the three corners, $P_{1}$, $P_{2}$, and $P_{3}$. Due to the
discreteness of this degree spectrum, it is convenient to work with
the cumulative distribution $P(k)=\sum_{k'\ge k} m(k',n)/N_{n}$,
where $N_{n}=3+(3^{n+1}-1)/2$ is the number of sites at generation
$n$.  In order to compute the analog to our problem of the degree
exponent $\gamma$, as commonly defined for continuum distributions
\cite{Doro2}, it is easy to show that $P(k) \propto k^{1-\gamma}$, 
with $\gamma = 1+\ln3/\ln2 \approx 2.585$.

\begin{figure}
\begin{center}
\includegraphics[width=0.4\textwidth]{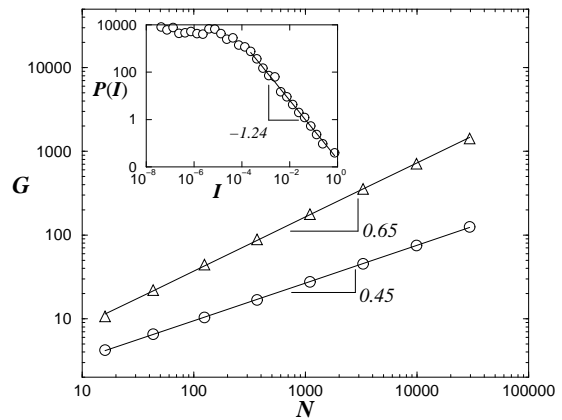}
\end{center}
\caption{ \label{Fig3} Conductance $G$ as function of the number
$N$ of sites when a potential is applied either between the center and
all the sites of the last generation (upper curve) or between two
corners of the network (lower curve). The inset shows the distribution
of local currents for the first case.}
\end{figure}

Next, we show that Apollonian networks display {\it small-world
effect} \cite{Watts,Amaral00,Doro3}. Accordingly, this means that the
average length of the shortest path $l$ between two vertices as
defined in ref.~\cite{Watts98} should grow slower than any positive
power of the system size $N$. For the network in Fig.~\ref{Fig1}(b) we
find that $l \propto (\ln N)^{\beta}$, with $\beta \approx 3/4$ as can
be seen in Fig.~\ref{Fig2}. This is a novel intermediate
behavior between {\it small} ($l \propto \ln N$) and {\it ultrasmall}
($l \propto \ln\ln N$) networks \cite{Rozenfeld02}.

The second property characterizing a small world network is the
clustering coefficient $C$ as defined in ref.~\cite{Watts98}. We find
it to be 0.828 in the limit of large $N$, as shown in the inset of
Fig.~\ref{Fig2}. This is a large value, comparable for instance
with the cluster coefficient for the collaboration network of movie
actors $C=0.79$ \cite{Watts98}. Consequently, since $L$ grows
logarithmically and $C$ is close to unity, the Apollonian network
indeed describes a small world scenario \cite{Watts98}.

As opposed to hierarchical lattices \cite{Doro2} used for instance
in the Migdal-Kadanoff approximation, Apollonian networks are
embedded in Euclidean space and can therefore describe geographical 
situations. In the case of Fig.~\ref{Fig1}(a) the sites form a 
multifractal point set in the asymptotic limit of infinitely
many iterations \cite{Manna91}. In the Apollonian networks for $n
\rightarrow \infty$, the bonds do not only completely cover the space
(like a Peano curve), but also never cross each other. In this situation
we have what is called a ``matching graph'' which is not the case for
any other known scale-free network \cite{Barabasi02}.

Now let us explore the properties of some well-known simple models
defined on the lattice shown in Fig.~\ref{Fig1}(b). Thinking of an
electrical supply system where the central site is the source (power
station), the last generation sites are sinks (consumers), and all the
connections have the same conductance, one can calculate the
equivalent conductance $G$ of the network. Going from one generation
to the next, the number of parallel system outputs increases by a
factor three while at each node the number of outputs doubles. Due to
the self-similarity, one has therefore doubled the conductance while
increasing the number of sites by three. As a result, $G$ follows a
power-law, $G \propto N^{z}$, with $z=2/3$. As shown in Fig.~\ref{Fig3}, 
this exponent can be confirmed through computational simulation of the
resistor network model. From Kirchhoff's law, the resulting system of
linear algebraic equations are numerically solved using a sparse
matrix inversion technique. This solution also allows us to compute
the distribution of local currents $P(I)$, as shown in the inset of
Fig.~\ref{Fig3}. Interestingly it follows a power-law for large values
of $I$ with an exponent $-1.24$, as a consequence of the hierarchical
choice of the output sites.

One can alternatively apply a potential drop between two corners of
the system, i.e. $P_1$ and $P_2$ in Fig.~\ref{Fig1}(b). The resulting
conductance numerically calculated as a function of the size of the
network gives a power-law too, but with exponent 0.45 as shown in
Fig.~\ref{Fig3}.

\begin{figure}
\begin{center}
\includegraphics[width=0.4\textwidth]{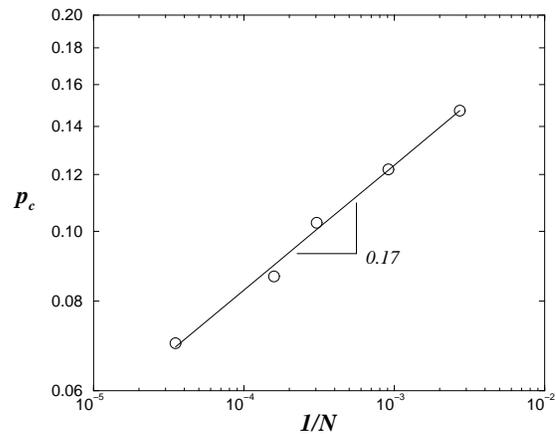}
\end{center}
\caption{\label{Fig4} Double logarithmic plot of the percolation
threshold $p_c$ as function of the inverse system size $N$.}
\end{figure}

We now proceed with the study of bond percolation simultaneously applied
among the three points $P_{1}, P_{2}$ and $P_3$ of Fig.~\ref{Fig1}(b) 
removing, however, the three direct connections between these
points. Similarly to ref.~\cite{Cohen00}, we find that the percolation
threshold $p_c$ goes to zero as the number of sites increases,
according to a power-law, as seen in Fig.~\ref{Fig4}. Assuming $p_{c}
\propto \sqrt{N}^{-\frac{1}{\nu}}$, the slope yields an exponent $\nu
\approx 3$.

Spin models can be defined on the Apollonian lattice in a variety of
ways. We consider here that an Ising spin $s_i$ is placed on each site
of the lattice. Between two sites we have a pair interaction constant
$J_n$, where $n$ indicates that the link between the interacting spins
is placed at the $n$-th generation of the construction of the lattice.
Moreover, the interactions are considered to be ferromagnetic $J_n>0$,
decaying with $n$ as $J_n \propto n^{-\alpha}$, and insensitive to the
actual distance between the spins. With this model definition, it is
straightforward to use a transfer matrix formalism to write down
temperature dependent recurrence relations for the free energy $f_n$
and correlation length $\xi_n$ of the model at two subsequent
generations, $f_{n+1}=f_{n+1}(f_n,\xi_n;T)$ and
$\xi_{n+1}=\xi_{n+1}(f_n,\xi_n;T)$ \cite{Kohmoto83,Andrade99}.  Exact
values for these and any other functions in the thermodynamical limit
$n\rightarrow\infty$ can be obtained through the numerical iteration
of $f_{n+1}$ and $\xi_{n+1}$ and for their derivatives. The iteration
process $n$ is stopped when the free energy per spin is the same in
two subsequent generations within the first 16 digits.

\begin{figure} [!h]
\begin{center}
\includegraphics[width=0.4\textwidth]{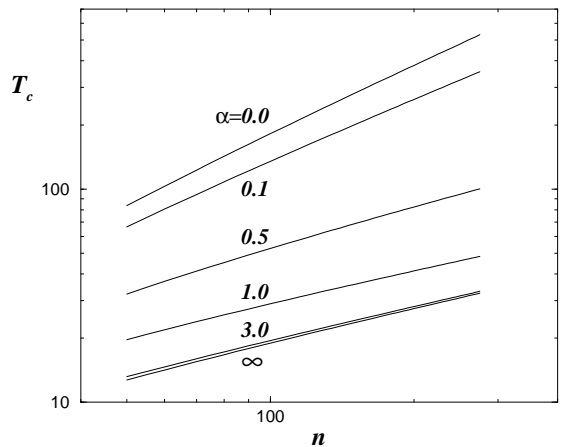}
\end{center}
\caption{\label{Fig5} Double logarithmic plot of the critical
temperature as function of the generation number $n$ for different
exponents $\alpha$.}
\end{figure}

The results obtained here are quite different from those for other
self-similar lattices. The free energy, entropy, specific heat and
magnetization (respectively $f,s,c,m$) are found to be completely
smooth for all values of $T$. A classical Schottky maximum dominates
the behavior of $c$ while $s$ increases monotonically to $k_B\ln2$. 
The presence of long-range order is observed as $m=1$ at low values of
$T$. On the other hand, when $T$ is large, $m\rightarrow 0$ as
$\exp(-T)$ for $\alpha=0$, with no noticeable presence of a critical
temperature. For $\alpha >0$, we find a stronger decay, as
$m\sim\exp(-T^\lambda)$. The results for the correlation length are
more subtle. $\xi$ diverges at low temperatures, expressing the
evidence of long-range order. As $T$ increases beyond a given $T_c(n)$
it converges to a well defined value. However, if the iteration
procedure is pursued to a value of $n$ larger than the one required
for the convergence of $f_n$, we observe that $\xi$ keeps on diverging
within a larger $T$ interval. This behavior therefore suggests the
existence of an $n$ dependent effective critical temperature $T_c(n)$,
as has been found by other authors for spin models on another
scale-free lattice \cite{Stauffer02}. In Fig.~\ref{Fig5} we show
how $T_c(n)$ depends on $n$ for different values of $\alpha$. The
power-law $T_c(n)\sim n^{\tau(\alpha)}$, with $\tau$ going
continuously from $\tau(0) = 1$ to $\tau(\infty) = 1/2$ indicates a
novel kind of critical behavior.

It is important to mention that a more general version of the network
shown in Fig.~\ref{Fig1}(b) is to choose at generation $n$ in each
triangle a new site randomly and connect it to the three adjacent
sites. In this sense the network can describe the porous fine
structure inside a block of heterogeneous soil surrounded by larger
pores or as the network of roads or supply lines going to smaller
urban units surrounded by larger ones. There exists in fact a very
large class of Apollonian networks obtained not only by changing the
choice of the position of the sites inside the areas or changing the
initial graph, but also having a different topology as classified in
ref.~\cite{Oron00}. All these networks have the properties presented
here, however, with different exponents if the topology is different.

Apollonian networks have the typical properties of the complex graphs
in nature and society: large clustering coefficients, slower than
logarithmic increase of the shortest path with number of sites,
and a power-law degree distribution. In the context of critical
phenomena, we did not find a finite critical point neither for
percolation nor for the Ising model on Apollonian networks. The
exponent $\nu = 3$ controlling how the percolation threshold goes to
zero with $N$ is unusually high while $T_c(n)$ of the Ising model
diverges with a power-law in the number $n$ of generations.

Considering the description of the force network in a polydisperse
packing, the small world nature of the lattice implies that between
any two grains there are only a few contacts at best. If the medium is
preconstrained, the electric conductance and elastic response have the
same divergence \cite{Alexander84}, and we find that the system becomes
more rigid with the inclusion of smaller particles following a power-law 
(with exponent 0.45 in our topology) in the number of particles. 
Giving to each grain a remnant magnetisation as discussed
in ref.~\cite{Roux93}, our result for the Ising model implies that, 
in the limit of many particles, one always find a ferromagnetic
response.

Interpreting our network as a porous medium, the fact that $p_c=0$ is
compatible with Archie's law, namely, that there exists no finite
critical porosity and the medium always percolates. If the network
represents roads or electric supply lines between cities implies that
they have small world character and that the distribution of car flux
or electrical currents follows a power-law when going from larger to
smaller sites. The results for the Ising model imply that opinions or
cooperations can stabilize at any degree of interaction, and the
percolation threshold being zero means that in the thermodynamic limit
the global connectivity cannot be disrupted, no matter how many
connections are broken.

We have seen that the Apollonian networks introduced in this paper can
have several applications yielding reasonable conclusions while on the
other hand their critical properties are rather special. They should
be studied in more detail including synchronization, spreading,
path-optimization, search algorithms and more specific features of the
real systems.

We thank CNPq, CAPES, and FUNCAP for support. We also thank Jason
Gallas for useful discussions.

\bibliographystyle{prsty}

\end{document}